\documentclass[icol,sn-nature,super]{sn-jnl}
\usepackage{graphicx}
\usepackage{multirow}%
\usepackage{amsmath,amssymb,amsfonts}%
\usepackage{amsthm}%
\usepackage{mathrsfs}%
\usepackage[title]{appendix}%
\usepackage{xcolor}%
\usepackage{textcomp}%
\usepackage{manyfoot}%
\usepackage{booktabs}%
\usepackage{algorithm}%
\usepackage{algorithmicx}%
\usepackage{algpseudocode}%
\usepackage{listings}%
\usepackage{anyfontsize}
\usepackage[version=4]{mhchem}
\usepackage{comment}
\usepackage{setspace}
\makeatletter
\long\def\@makecaption#1#2{%
  \vskip\abovecaptionskip
  \normalsize 
  \textbf{#1.} #2\par
  \vskip\belowcaptionskip}
\makeatother
\newcommand{\cofirstdag}{$^{\dagger}$}
\begin{document}

\title{\textbf{Electrically switchable ferron upconversion in a van der Waals ferroelectric}}
\author[1]{\fnm{Sujan} \sur{Subedi}\cofirstdag}
\author[1]{\fnm{Wuzhang} \sur{Fang}\cofirstdag}
\author[1]{\fnm{Fan} \sur{Fei}\cofirstdag}
\author[2]{\fnm{Zixin} \sur{Zhai}\cofirstdag}
\author[1]{\fnm{Jack P.} \sur{Rollins}}

\author[1]{\fnm{Carter} \sur{Fox}}
\author[1]{\fnm{Alaina} \sur{Drew}}
\author*[2]{\fnm{Bing} \sur{Lv}} \email{blv@utdallas.edu}

\author*[1,3,4]{\fnm{Yuan} \sur{Ping}}\email{yping3@wisc.edu}
\author*[1,3,5]{\fnm{Jun} \sur{Xiao}}\email{jun.xiao@wisc.edu}

\affil[1]{\orgdiv{Department of Materials Science and Engineering}, \orgname{University of Wisconsin-Madison}\orgaddress{\street{}, \city{Madison}, \postcode{53706}, \state{Wisconsin}, \country{USA}}}

\affil[2]{\orgdiv{Department of Physics}, \orgname{University of Texas at Dallas}\orgaddress{\street{}, \city{Richardson}, \postcode{75080}, \state{Texas}, \country{USA}}}
\affil[3]{\orgdiv{Department of Physics}, \orgname{University of Wisconsin-Madison}\orgaddress{\street{}, \city{Madison}, \postcode{53706}, \state{Wisconsin}, \country{USA}}}

\affil[4]{\orgdiv{Department of Chemistry}, \orgname{University of Wisconsin-Madison}\orgaddress{\street{}, \city{Madison}, \postcode{53706}, \state{Wisconsin}, \country{USA}}}

\affil[5]{\orgdiv{Department of Electrical and Computer Engineering}, \orgname{University of Wisconsin-Madison}\orgaddress{\street{}, \city{Madison}, \postcode{53706}, \state{Wisconsin}, \country{USA}}}

\affil[$^{\dagger}$]{These authors contributed equally to this work}

\maketitle
\section*{Abstract}

\textbf{Nonlinear phononics provides a powerful ultrafast route to control lattice excitations, enabling access to hidden quantum orders, phononic computing, and quantum transduction. However, dynamic control of anharmonic phonon interactions remains limited, as these interactions are typically fixed by the equilibrium crystal lattice and lack external tunability. Emergent ferrons in ferroelectrics, which are collective oscillations of the spontaneous electric polarization, may offer a promising platform to overcome this limitation by combining intrinsic phononic nonlinearity with direct electrical control of the ferroelectric order parameter. Here we report electrically controllable nonlinear ferron upconversion in the van der Waals ferroelectric NbOI\textsubscript{2}. We show that resonant THz excitation of a 3.1 THz ferron drives coherent upconversion to a 7.0 THz optical phonon. Using two-dimensional THz spectroscopy, we directly resolve off-diagonal coupling features and establish the nonlinear upconversion pathway. Supported by first-principles calculations and analytical modeling, we identify the microscopic origin as a cubic anharmonic lattice coupling. Importantly, in situ electric-field switching enables nonvolatile control of both the ferron dynamics and the associated upconversion process. The phase reversal and hysteretic behavior across the coercive fields establish that the ferron-mediated nonlinear phononic interaction is strongly dependent on the underlying ferroelectric order parameter. These results introduce ferron upconversion as a new and universal regime of nonlinear phononics in ferroelectrics and establish an electrically programmable platform for coherent lattice control, paving the way for ferronic information processing and quantum phononic transduction.
}


\maketitle
 
 \newpage

\section*{Main}

Driving solids far from equilibrium enables access to collective dynamics that are inaccessible under linear response conditions. In nonlinear phononics, resonant excitation of lattice vibrations into the anharmonic regime allows transient symmetry modification and free energy renormalization for revealing novel quantum orders, including light-driven ferromagnetim in antiferromagnets, light-induced high temperature superconductivity, ultrafast manipulation of ferroelectricity and photon-induced chirality in a nonchiral crystal \cite{disa2020polarizing,disa2021engineering,nova2017effective,stupakiewicz2021ultrafast,zeng2025photo,mitrano2016possible,minakova2025direct,mankowsky2017ultrafast,li2019terahertz}. Beyond uncovering hidden phases, nonlinear phononic coupling governs coherent interactions among lattice quantum excitations and provides a conceptual foundation for phononic logic operations, hybrid quantum transduction, and phononic quantum entanglement \cite{pirie2022topological,neuman2021phononic,chou2025deterministic}. Despite the great potential,  achieving on-demand control of nonlinear phononic processes remains a major challenge in this research frontier \cite{disa2021engineering}. In most current nonlinear phononics studies, anharmonic phonon interactions are constrained by the equilibrium lattice configuration and lack external tunability. This constraint limits full access to hidden phase diagram and prevents programmable control of coherent phononic coupling dynamics.

A promising route to overcome this limitation lies in ferroelectric quantum materials, where the spontaneous polarization acts as a macroscopic order parameter and is directly coupled and switchable by electric fields. Its collective excitations, known as ferrons, have recently been predicted as coherent oscillations of the ferroelectric polarization \cite{tang2024electric,tang2022excitations,bauer2023polarization}. This excitation mode enables direct polarization transport without charge flow, carrying information encoded in its amplitude and phase. For instance, coherent ferrons have recently been discovered in van der Waals ferroelectrics NbOI\textsubscript{2} with a sharply defined mode near 3.1 THz, ultralong lifetimes exceeding 100\,ps  \cite{huang2025coupling} and hypersonic propagation velocity over 10\textsuperscript{5}\,m/s \cite{choe2025observation}. The combination of long lifetime and high group velocity yields coherent propagation lengths of approximately 20\,$\mathrm{\mu m}$, comparable to their magnon counterparts in 3D and 2D magnets such as YIG\cite{liu2018long}, CrSBr \cite{bae2022exciton} and MnPS\textsubscript{3} \cite{alliati2022relativistic}.  More importantly, owing to their intrinsic dipolar character, ferrons are expected to couple strongly to electromagnetic fields and exhibit pronounced nonlinearity and electrical controllability, offering a pathway toward dynamically reconfigurable phononic interactions. Despite these advances, ferron dynamics have thus far been confined to the linear regime. A key fundamental question is whether the polarization mode itself can mediate nonlinear coupling between lattice excitations and thereby extend nonlinear phononics into a polarization order parameter governed and electrically reconfigurable regime. 

Here, we demonstrate THz driven ferron upconversion with nonvolatile electrical field control in the layered van der Waals ferroelectric NbOI\textsubscript{2}. Using resonant THz excitation and 2D THz spectroscopy, we directly resolve nonlinear coupling between the low frequency ferron mode and a longitudinal optical phonon mode. 
We further show that both the ferron dynamics and the nonlinear upconversion response are dynamically tunable by electric field switching. Our results establish ferron mediated upconversion as a previously unexplored nonlinear regime of lattice dynamics and highlight ferrons as new active participants in nonlinear phononics with their unique dipolar nature and electric field tunability.

\section*{Coherent excitation of THz ferron and other optical phonon modes }

\begin{figure*}[htb]
     \centering \includegraphics[width=\linewidth]{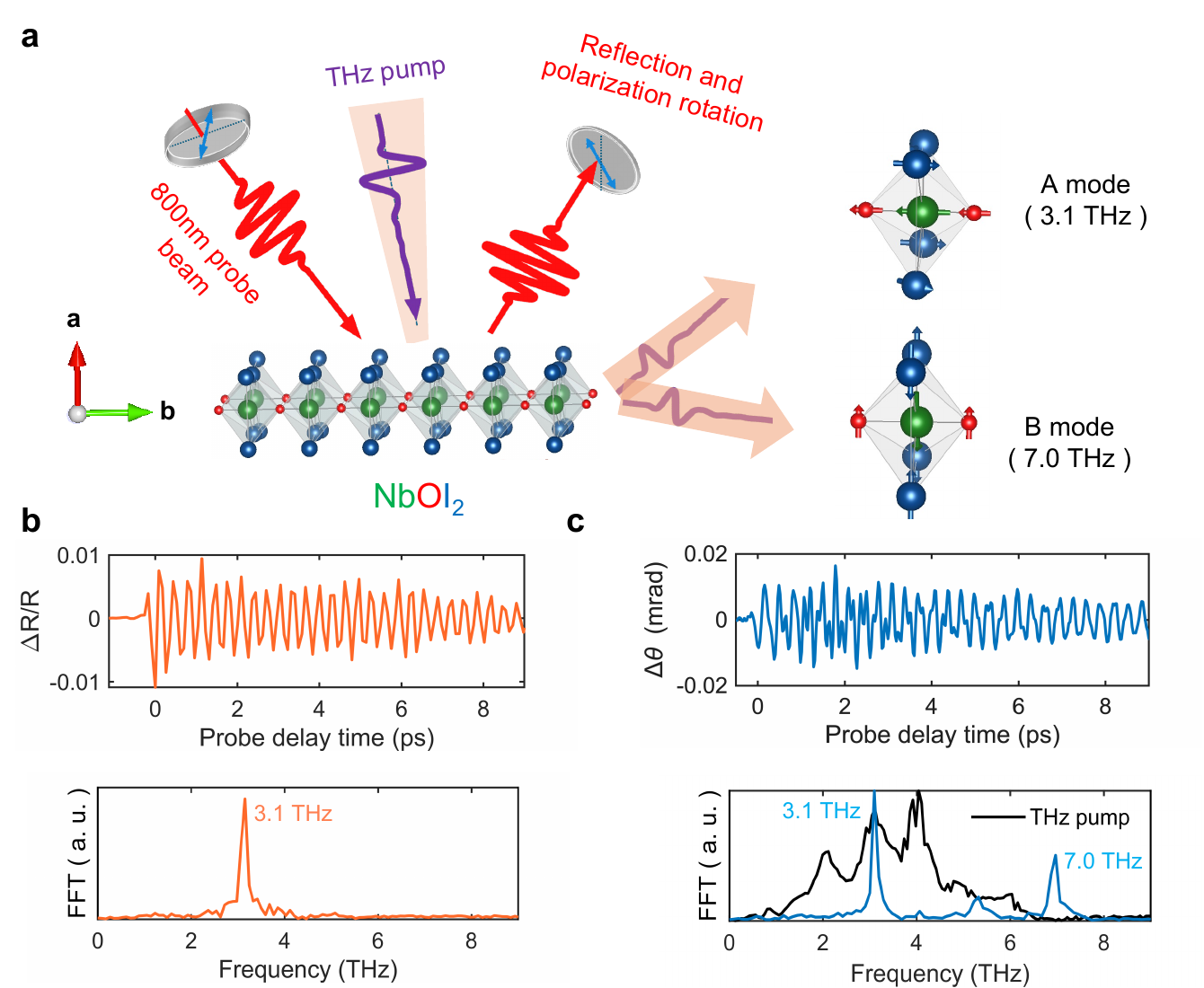}
     \caption{
     Coherent THz excitation of ferron and optical phonon modes in vdW ferroelectric NbOI\textsubscript{2}. (a) Schematic of the THz pump–optical probe experiment used to phononic excitations in NbOI\textsubscript{2}, with Nb (green), I (blue), and O (red). Arrows indicate the calculated atomic displacement patterns of the A-symmetry 3.1 THz ferron mode and the B-symmetry 7.0 THz optical phonon mode. (b) Time-resolved optical reflectivity change under THz excitation with the field aligned along the ferroelectric axis, revealing coherent oscillations of the ferron mode at 3.1 THz. (c) Time-resolved polarization rotation signal showing coherent oscillations at both the ferron mode (3.1 THz) and a higher-frequency optical phonon at 7.0 THz. The latter lies outside the spectral bandwidth of the THz pump, indicating a nonlinear excitation pathway.}\label{Main_fig:1}
\end{figure*}
NbOI\textsubscript{2} is a layered van der Waals ferroelectric crystallizing in a noncentrosymmetric monoclinic C2 structure. In this structure, niobium cations exhibit off-center displacements along the Nb–O–Nb atomic chain direction, resulting in a large spontaneous polarization of approximately $20\,\mu\mathrm{C}/\mathrm{cm}^2$, among the highest reported for vdW ferroelectrics \cite{zhang2024giant,jia2019niobium}.
The polar space group and large ferroelectric polarization supports a ferron mode, namely a polar optical phonon associated with a time-varing ferroelectric polarization \cite{subedi2025colossal,huang2025coupling}. Its polar optical phonon nature dominated by the short-range interatomic interactions, results in ferron frequencies in the THz regime, which offers ultrafast information processing beyond the GHz limits of conventional magnons. On the other hand, the long-range dipolar interactions among electric dipoles can yield high group velocity to 10\textsuperscript{5} m/s, by inducing steeper dispersion, promoting phase correlation and smoothing local inhomogeneities \cite{choe2025observation}. Besides this unique THz ferron mode, the first principle calculations also suggest several THz Raman- and infrared-active phonon modes nearby (see {\bf Supplementary Section S1 Table T1}). The coexistence of multiple polar and nonpolar optical phonon modes in THz regimes makes NbOI\textsubscript{2} as a promising platform for efficient nonlinear phononic coupling.

To investigate the possible interplay between different THz phonon modes,  we first prepared thick and large NbOI\textsubscript{2} flakes and pumped them by intense single-cycle THz pulses generated via optical rectification in an organic nonlinear crystal (See {\bf Methods} and {\bf Supplimentary Section S2} for more information).  Distinct from previous optical excitation studies, resonant THz driving maximizes the ferron oscillation amplitude and enables access to nonlinear phonon coupling that scales nonlinearly with the driving field. The peak electric field at the sample reached approximately 770\,kV/cm, calibrated by typical electro-optic sampling method \cite{blank2023two,blanchard2007generation} (see {\bf Supplementary Section S3} for details). To probe the THz-driven phononic dynamics, a weak 800\,nm linearly polarized pulse was co-propagated with the THz pump to monitor its transient changes in reflectivity and polarization rotation, respectively.  As shown in Fig. \ref{Main_fig:1}b, time-resolved reflectivity measurements reveal a narrowband oscillation at 3.1 THz, consistent with the long-lived ferron mode previously reported in NbOI\textsubscript{2}\cite{choe2025observation,handa2025terahertz}. Besides, our time-resolved second harmonic generation measurements further confirm that this oscillation corresponds to coherent polarization dynamics at the same frequency, establishing its polarization wave origin (see {\bf Supplementary Section S4}). More interestingly, time-resolved optical polarization rotation measurements resolve dynamic signatures of additional collective modes. For example, with the probe polarization aligned along the polar $b$-axis,  we observed two major phononic excitations with frequencies at 3.1\,THz and 7.0\,THz, along with two minor oscillation modes at 4.1\,THz and 5.3\,THz ( Figure \ref{Main_fig:1}c). The 3.1 THz mode is symmetric with respect to the polar \textit{b}-axis and corresponds to an A-symmetry phonon, whereas the 7.0 THz mode involves atomic motion along the nonpolar direction and is antisymmetric with respect to the polar axis, consistent with B-symmetry (Figure \ref{Main_fig:1}a). Consistent with these symmetries, the 3.1 THz mode exhibits transverse optical (TO) character with atomic displacements along the polar \textit{b}-axis, whereas the 7.0 THz mode exhibits longitudinal optical (LO) character with atomic motions parallel to the wave propagation direction. Although the 7.0 THz mode corresponds predominantly to an out-of-plane vibration, it exhibits a finite in-plane anisotropic dielectric response. As a result, this mode primarily induces birefringence with minimal change in the diagonal refractive index, rendering it nearly invisible in reflectivity measurements while remaining clearly detectable through polarization rotation (See {\bf Supplementary Section S1} for more details).  
Regarding the origin of this excitation, the terahertz pump spectrum extends only up to approximately 6.1 THz, allowing direct resonant excitation of the 3.1, 4.1, and 5.3 THz infrared-active modes. The 7.0 THz mode, however, lies well beyond the pump bandwidth and therefore cannot be directly driven by the incident terahertz field. To further verify this, we applied THz filters to restrict the pump spectrum below 3.5 THz. Under these conditions, the 7.0 THz oscillation remains clearly observable (see {\bf Supplementary Section S5} for more details). Moreover, this mode cannot originate from sum-frequency generation between the 3.1 and 4.1 THz modes, because such a nonlinear mixing process requires the simultaneous excitation of both modes, while the THz filter supresses the the direct excitation of the 4.1 THz mode. The persistence of this higher-frequency response indicates that it may arise from nonlinear phonon coupling to the 3.1 THz ferron mode.

\section*{Polarization and fluence dependence of THz-driven phonon excitations }

\begin{figure*}[htb]
     \centering   \includegraphics[width=\linewidth]{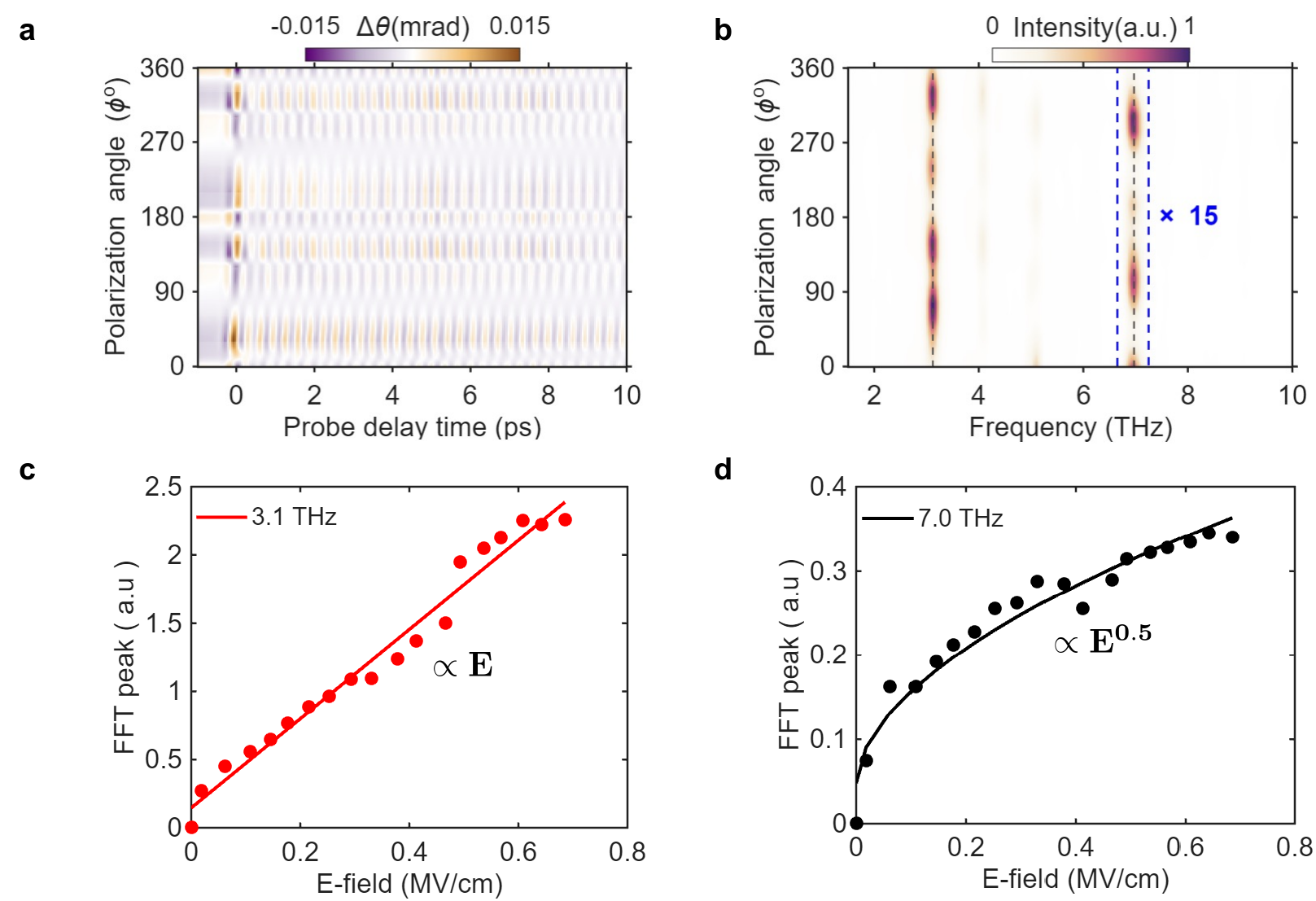}
     \caption{
     Polarization and fluence dependence of THz-driven coherent phonon excitations in NbOI\textsubscript{2}. (a) Polarization-resolved time-domain traces of the THz-induced optical polarization rotation measured for different probe polarization angles relative to the crystal axes. (b) Extracted amplitude of the ferron mode (3.1\,THz) and higher frequency mode (7.0\,THz) as a function of probe polarization angle, revealing distinct symmetry-dependent detection responses. (c,d) THz field dependence of the Fourier-transformed peak amplitudes for the (c) 3.1\,THz ferron mode and (d) 7.0\,THz optical phonon mode. The ferron mode exhibits approximately linear scaling with field ($\gamma\approx 1$), consistent with direct resonant excitation. The 7.0\,THz mode exhibits a sublinear dependence indicative of nonlinear phononic coupling.}\label{Main_fig:2}
\end{figure*}

To elucidate the symmetry selection and nonlinear coupling mechanisms for this THz ferron upconversion process,  THz-driven phononic excitations in NbOI\textsubscript{2}, we performed polarization-resolved and field-dependent THz pump–probe optical polarization-rotation measurements. In the polarization-resolved configuration, the incident polarization of the 800 nm probe was rotated while the linearly polarized THz pump field was fixed along the polar $b$-axis ($\theta=0^{\circ}$). This geometry enables selective detection of anisotropic optical responses arising from transient birefringence induced by different phonon modes. As shown in Fig. \ref{Main_fig:2}a, the oscillatory response is dominated by the 3.1 THz ferron mode, accompanied by additional higher-frequency components that become more pronounced at particular probe orientations. This behavior becomes clearer in the frequency domain after applying FFT to the time-domain data in  Fig. \ref{Main_fig:2}a. Figure \ref{Main_fig:2}b quantifies this anisotropy by extracting the angular dependence of the oscillation amplitudes. The ferron mode exhibits a pronounced four-lobed pattern with maxima near 45° relative to the polar axis and strong suppression along the principal crystal axes. The suppression of the ferron mode along the crystalline axes, with its maximum response at $45^{\circ}$ relative to the polar axis, is consistent with a signal arising from in-plane linear birefringence induced by the anisotropic dielectric response. In contrast, the 7.0\, THz mode displays a distinct angular response. It becomes most visible when the ferron contribution is suppressed near the crystalline axes, where the ferron-induced birefringence vanishes. The different angular dependence reflects the distinct symmetry of the phonon eigenvectors: the 3.1 THz ferron corresponds to an A-type vibration, whereas the 7.0 THz mode belongs to a B-type symmetry. Despite their different detection symmetries, both modes exhibit the same pump-angle dependence (see {\bf Supplementary Section S6}), indicating that the symmetry difference originates from the optical detection process rather than from the THz driving mechanism.\\

To further examine the excitation mechanism, we measured the transient optical polarization rotation signals as a function of the incident THz pump electric-field amplitude $E_{\mathrm{THz}}$ (Fig. \ref{Main_fig:2}c) (see {\bf Supplementary Section S7} for the detail).  The amplitudes of the individual modes were obtained by extracting the peak values of the Fourier-transformed oscillatory signals. The resulting field dependence is summarized in Fig. \ref{Main_fig:2}d.  In particular, the 3.1\,THz ferron amplitude increases linearly with the pump field ($\gamma \approx 1$), consistent with direct resonant excitation origin. In contrast, the 7.0\,THz mode exhibits a sublinear dependence ($\gamma \approx 0.5$), which cannot be explained by direct driving and instead points to a nonlinear excitation pathway. Such behavior is consistent with anharmonic phonon coupling in which the resonantly driven ferron mediates the generation of the higher-frequency phonon.

\section*{Observation of ferron upconversion by two-dimensional THz spectroscopy}

\begin{figure*}[htb]
     \centering     \includegraphics[width=0.96\textwidth]{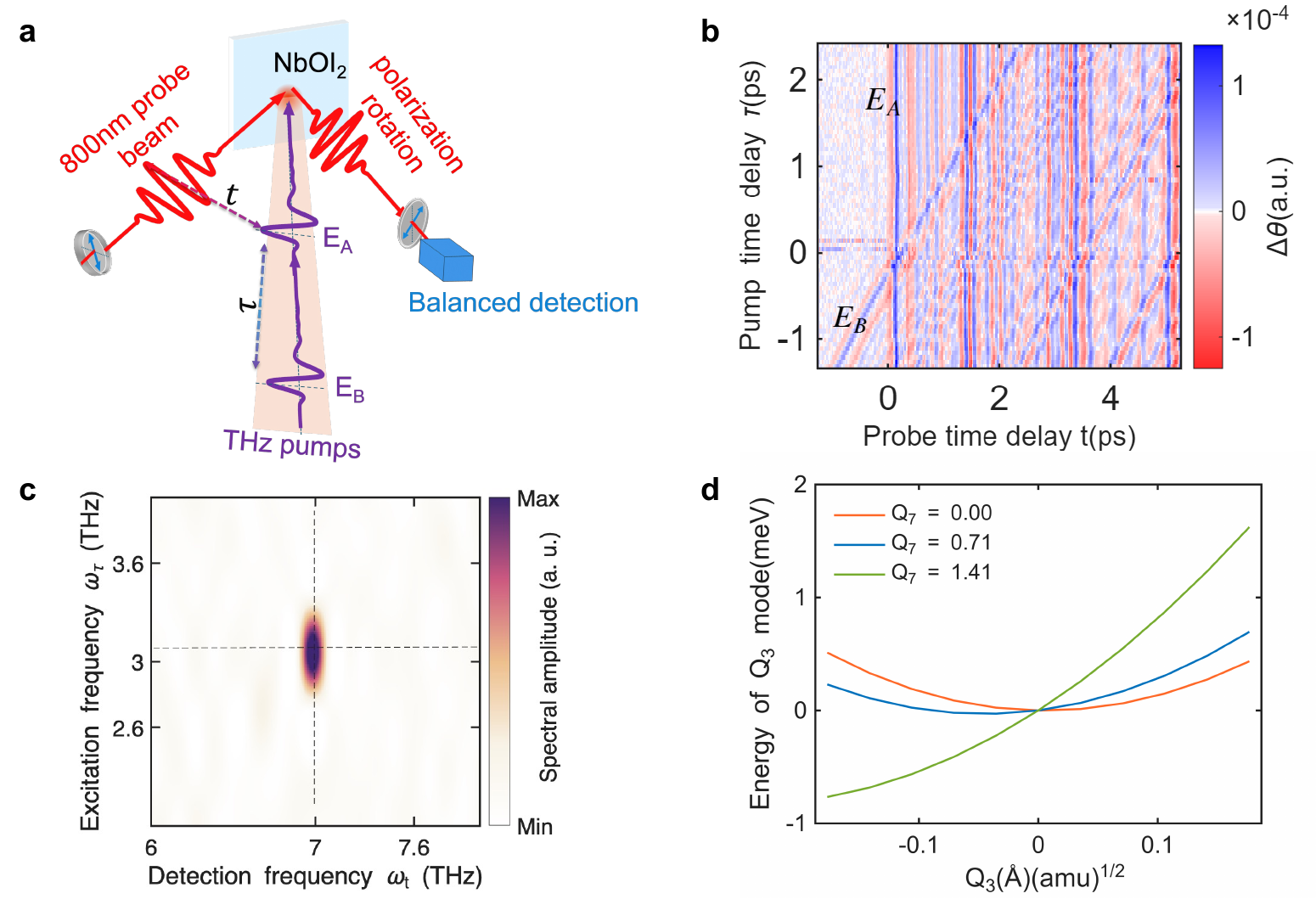}
      \caption{Two-dimensional THz spectroscopy revealing the ferron and longitudinal optical phonon coupling in NbOI\textsubscript{2}. (a) Schematic of the two-dimensional THz spectroscopy. Two single-cycle THz pump pulses generated via optical rectification in a DSTMS crystal are separated by an inter-pump delay $\tau$ and and focused collinearly onto the NbOI\textsubscript{2} sample. The induced polarization rotation of an 800 nm probe pulse is recorded as a function of the probe delay time $t$ and inter-pump delay $\tau$. (b) Double THz pump-induced optical polarization rotation plotted as a function of  $t$ and $\tau$. (c) Two-dimensional fast Fourier transform (2D FFT) of the nonlinear signal as a function of detection frequency $f_\text{det}$ and excitation frequency $f_\text{exc}$. A pronounced off-diagonal feature appears at $f_\text{det}= 7.0$\,THz when the sample is driven at $f_\text{exc}= 3.1$\,THz, unambiguously revealing nonlinear coupling between the ferron and the higher frequency optical phonon.  (d) Calculated lattice energy landscape as a function of the ferron mode amplitude $Q_3$ for several values of the higher frequency optical phonon mode amplitude $Q_7$. The energy for zero displacement is taken as the reference point.} \label{Main_fig:3}
\end{figure*}

To unambiguously establish the nonlinear phonon coupling between the 3.1 THz ferron mode and the 7.0 THz phonon mode, we further performed two-dimensional terahertz (2D-THz) spectroscopy on NbOI\textsubscript{2} flakes. This technique provides a direct approach for resolving energy transfer and coherent coupling between phonon modes \cite{blank2023two,grishunin2023two,lin2022mapping}. In the resulting two-dimensional spectra, diagonal features correspond to linear phonon resonances, whereas off-diagonal peaks provide clear signatures of nonlinear coupling between different collective excitations \cite{liu2025multidimensional}. Our 2D-THz experimental configuration is illustrated in Fig.~\ref{Main_fig:3}a. Two single-cycle THz pump pulses (A and B) with a controllable inter-pulse delay ($\tau$) were focused collinearly onto the NbOI\textsubscript{2} sample. The induced optical polarization rotation was monitored using another 800 nm probe pulse and detected with a balanced detection scheme. By scanning both the probe delay time $t$ and inter-pump delay $\tau$, we obtained a two-dimensional time-domain dataset that captures the evolution of phonon coherences under multi-pulse excitation (see {\bf Methods} and {\bf Supplementary S8} for experimental details).

To isolate the nonlinear response arising from coherent interactions between the two THz pulses, we adopted a stepwise measurement protocol \cite{woerner2021two,raab2019ultrafast,mukamel1995principles}. First, the response to pulse A alone, $S_{\mathrm{A}}(t,\tau)$, was measured by blocking pulse B. Next, the response to pulse B alone, $S_{\mathrm{B}}(t,\tau)$, was recorded by blocking pulse A. Finally, both pulses were applied simultaneously to measure the total response, $S_{\mathrm{AB}}(t,\tau)$, shown in Fig. \ref{Main_fig:3}(b). The nonlinear signal $S_{\mathrm{NL}}(t,\tau)$ was then extracted using
\begin{equation}
S_{\mathrm{NL}}(t,\tau) = S_{AB}(t,\tau) - S_{A}(t,\tau) - S_{B}(t,\tau)
\end{equation}
This subtraction removes all linear and additive contributions, leaving only the nonlinear response associated with coherent multi-pulse interactions and intermode energy transfer processes \cite{tian2003femtosecond,lu2018two,kuehn2011two,reimann2021two}. Then we applied a two-dimensional fast Fourier transform (2D FFT) to $S_{\mathrm{NL}}(t,\tau)$, producing the frequency–frequency spectrum shown in Fig. \ref{Main_fig:3}c. The resulting map is plotted as $\theta_\text{NL}(f_\text{det},f_\text{exc})$, where $f_\text{det}$ and $f_\text{exc}$ represent the detection and excitation frequencies, respectively. The spectrum reveals a pronounced off-diagonal peak linking the the ferron excitation at ${f}_\text{exc} = 3.1$\,THz, with the higher frequency phonon mode at $f_{\text{det}} = 7.0$\,THz. Such off-diagonal resonances are a hallmark of nonlinear phonon coupling and provide direct evidence that the resonantly driven ferron coherently generates the higher-frequency phonon through an anharmonic lattice interaction. \cite{blank2023two,mashkovich2021terahertz,zhang2024terahertz,johnson2019distinguishing}

To elucidate the microscopic origin of this ferron-mediated upconversion process, we model the THz-driven lattice dynamics using an anharmonic oscillator framework involving the ferron coordinate $Q_3$ (3.1 THz), and the higher-frequency phonon coordinate $Q_7$ (7.0\,THz). The symmetry-allowed lattice potential consistent with the C2 crystal symmetry, in terms of the mode $Q_3$ and $Q_7$ up to the fourth order, can be expressed as
\begin{equation}
V(Q_{3},Q_{7})
=Z^{*} Q_{3} E-
g\,Q_{3}Q_{7}^{2}
-
d\,Q_{3}^{2}Q_{7}^{2}
\end{equation}
where $Z^*$  is effective mode charge, \textit{E} is applied THz electric field, \(g\)  and \(d\)  are nonlinear coupling coefficients. Because the $Q_7$ phonon corresponds to a lattice vibration along the nonpolar axis, symmetry requires the free energy to contain only even powers of $Q_7$ . First-principles calculations further support this coupling mechanism. By computing the lattice energy landscape as a function of  $Q_3$ for several fixed values of $Q_7$ (Fig \ref{Main_fig:3}d), we find that a finite $Q_7$ displacement shifts the equilibrium position of the $Q_3$ coordinate and produces a pronounced asymmetric energy profile. This behavior rules out symmetric coupling terms such as $Q_3^2Q_7^{2}$ and identifies the $Q_3 Q_7^{2}$ interactions as the dominant nonlinear coupling channel.  Using experimentally determined damping parameters and fitting the model to the measured nonlinear dynamics, we extract a coupling constant $g$ of 8.4\,$\mathrm{meV}\,\text{\AA}\,\mathrm{amu}^{-3/2}$ (see {\bf Supplementary S9} for the method of calculation). This value is consistent with first-principles calculation estimates of 3.34\,$\mathrm{meV}\,\text{\AA}\,\mathrm{amu}^{-3/2}$. It is also comparable to cubic anharmonic coupling strength reported in orthoferrites such as ErFeO\textsubscript{3} \cite{juraschek2017ultrafast} and in multiferroics such as BiFeO\textsubscript{3} \cite{bustamante2025ultrafast}.  Together, these results quantitatively confirm that the observed 7.0 THz oscillation arises from nonlinear upconversion mediated by the resonantly driven ferron mode.

\section*{In situ electric-field control of ferron upconversion}

\begin{figure*}[htb]
     \centering   \includegraphics[width=\linewidth]{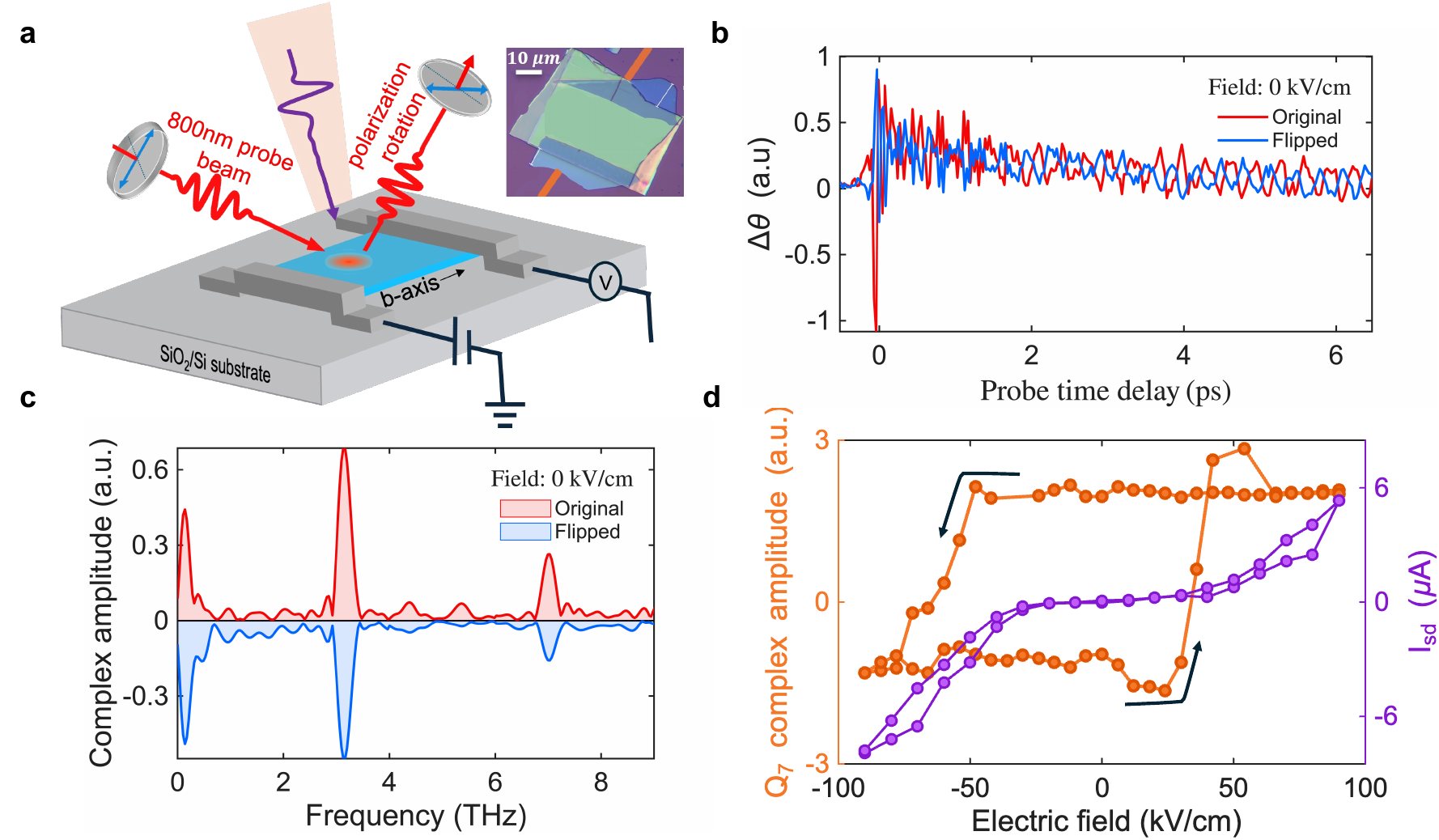}
     \caption{In situ electric-field control of ferron upconversion in NbOI\textsubscript{2}. (a) Schematic of the device used to apply a static electric field to the NbOI\textsubscript{2} flake. The inset shows an optical image of the device with graphene electrodes separated by 25\,$\mu m$. The electric field is applied along the polar \textit{b}-axis. (b)  Time-domain traces of the THz pump-induced optical polarization rotation change ($\Delta\theta$) measured at 0 kV/cm before and after electrical polarization switching. Reversing the ferroelectric polarization leads to a clear inversion of the ferron oscillation phase. (c) Fourier-transformed spectra of the ferron at 3.1 THz and the optical phonon at 7.0\,THz at 0\,kV/cm before and after polarization reversal. (d) Electric-field dependence of complex amplitude of upconverted 7 THz mode (left axis), together with the simultaneously recorded source-drain current I$\textsubscript{sd}$ (right axis). The sweep direction is indicated by the black arrows. The complex amplitude exhibits a hysteretic response that follows the ferroelectric polarization switching, evidencing nonvolatile electrical control of the ferron upconversion process.} \label{Main_fig:4}
\end{figure*}

Finally, we explore the nonvolatile electrical control of ferron dynamics and the associated nonlinear phonon coupling. To directly demonstrate the control, we fabricated in-plane two-terminal NbOI\textsubscript{2} devices and performed in situ electric-field dependent THz pump-optical probe measurements (Figure \ref{Main_fig:4}a). The device consists of a thick NbOI\textsubscript{2} flake (approximately 150\,$\mathrm{nm}$ thick and $\sim 65 \times 90\,\mu\mathrm{m}$ in lateral size) exfoliated from a bulk crystal and transferred onto a silicon substrate patterned with two graphene electrodes separated by about 25\,$\mu$m. The ferroelectric polar axis was aligned along the applied static electric field. The ferroelectric polar axis of the crystal was aligned along the direction of the applied static electric field. Time-resolved polarization rotation signals were recorded while sweeping the applied electric field between $-90$\,kV\,cm$^{-1}$ to $+90$\,kV\,cm$^{-1}$. To prepare the initial polarization states, a field of $-90$\,kV\,cm$^{-1}$ was first applied to pole the ferroelectric polarization, after which the field was returned to 0\,kV\,cm$^{-1}$ for measurement. The field was then swept to $+90$\,kV\,cm$^{-1}$ to reverse the polarization and subsequently returned again to 0\,kV\,cm$^{-1}$. As a result, the two measurements conducted at 0\,kV\,cm$^{-1}$ correspond to opposite remanent polarization states of the ferroelectric crystal.

Fig.~\ref{Main_fig:4}b shows representative time-domain polarization-rotation traces measured at zero field before and after polarization switching, with the corresponding Fourier spectra presented in Fig.~\ref{Main_fig:4}c. Both the ferron mode at 3.1 THz and the upconverted 7.0 THz phonon exhibit a clear phase reversal between the two polarization states. This sign change directly reflects the reversal of the ferroelectric polarization and demonstrates that both the ferron oscillation and its nonlinear upconversion are governed by the macroscopic ferroelectric order parameter. Small differences in oscillation amplitude between the two states likely arise from incomplete polarization reversal across domains, which may lead to partial signal cancellation. To further establish the connection between the upconverted 7 THz optical phonon response and ferroelectric switching, we continuously swept the electric field between $\pm 90$\,kVcm$^{-1}$, while simultaneously monitoring the FFT complex amplitude of upconverted 7.0 THz mode and the source-drain current (I$\textsubscript{sd}$). As shown in Fig.~\ref{Main_fig:4}d, both signals exhibit pronounced hysteresis loops characteristic associated with ferroelectric domain switching. The close correspondence between the optical polarization response and the electrical transport signal demonstrates that the coherent oscillations are governed by the remanent ferroelectric polarization state. These results provide direct spectroscopic evidence that both ferron dynamics and ferron-mediated upconversion can be controlled by an external electric field and exhibit nonvolatile behavior.

\section*{Conclusion}\label{sec13}

In summary, we have discovered electrically switchable ferron upconversion in the van der Waals ferroelectric NbOI$_2$, revealing a nonlinear phonon interaction between the 3.1\,THz ferron mode and the 7.0\,THz optical phonon mode. Two-dimensional THz spectroscopy directly resolves this nonlinear coupling through off-diagonal spectral features, while field-dependent scaling and phase reversal confirm its anharmonic origin. Combined with first-principles calculations and free-energy modeling, we identify a dominant  {$Q_3Q_7^{2}$} nonlinear phononic interaction and extract an effective coupling strength of approximately $\sim$8.4~meV~\AA~amu$^{-3/2}$, comparable to cubic anharmonic coupling reported in orthoferrites such as ErFeO\textsubscript{3} \cite{juraschek2017ultrafast} and multiferroics such as BiFeO\textsubscript{3} \cite{bustamante2025ultrafast}. Importantly, in situ electric-field switching establishes nonvolatile control of both the ferron dynamics and the associated upconversion process. The phase reversal and hysteretic behavior of the coherent ferron upconversion across the coercive fields demonstrate that the ferron-mediated nonlinear phononic interaction is strongly dependent on the macroscopic ferroelectric order parameter.

These results establish ferrons as highly active dipolar quasiparticles in the terahertz regime for nonlinear phononics. Unlike conventional nonlinear phonon processes that are fixed by lattice symmetry, the mechanism demonstrated here is intrinsically tied to a switchable broken symmetry and benefits from the strong electric-dipole character of the polarization wave. In NbOI\textsubscript{2}, the coexistence of a long-lived THz ferron mode, strong dipolar interactions, and closely spaced optical phonons provides a unique platform for electrically reconfigurable coherent lattice dynamics. In this regime, nonlinear intermode interactions become programmable through ferroelectric polarization switching, enabling nonvolatile control of polarization-wave transport and nonlinear coupling pathways. Beyond classical nonlinear phononics, the identified anharmonic interaction provides a key ingredient for spontaneous parametric down-conversion of THz phonons \cite{LibbiKozinsky2025QuantumNonlinearPhononics,Liu2013OptomechPDC,Wang2025NonclassicalPhononPair}, offering a pathway to generate correlated or entangled phonon pairs from a driven ferron mode. By linking nonlinear phonon conversion to a switchable ferroelectric order parameter, electrically programmable ferron upconversion also establishes a platform for reconfigurable THz collective dynamics and ferronic information processing in low-dimensional quantum materials, defines a ferronic analogue of nonlinear magnonics \cite{Zheng2023NonlinearMagnonics,Bracher2017ParallelPumping,Wang2016MagnonKerr,Wang2018MagnonPolaritonBistability}.

\section*{Methods}
\bmhead{Crystal growth and prepration of NbOI\textsubscript{2}}
NbOI\textsubscript{2} single crystals were synthesized using the chemical vapor transport method.  Niobium powder (99.8\%, Alfa Aesar), Nb\textsubscript{2}O\textsubscript{5} powder (99.9\%, Alfa Aesar) and Iodine flakes (99.99\%, Alfa Aesar) were mixed in an appropriate ratio under an inert atmosphere. The mixtures were loaded into a silica tube, which was flame-sealed under vacuum. The ampoule was then loaded in a horizontal tube furnace with the hot end containing the reactants maintained at 700\,$^\circ$C and the cold end at 650\,$^\circ$C. After 10 days of reaction, plate-like single crystals with size of several millimeters formed at the cold end. Bulk crystals were first cleaved repeatedly with adhesive tape to isolate thin layers. The exfoliated material was then gently pressed onto a polydimethylsiloxane (PDMS) stamp mounted on a clean glass slide. Upon peeling the tape slowly, individual flakes adhered to the PDMS surface. The PDMS was brought into gentle contact with a SiO\textsubscript{2}/Si substrate, held briefly to ensure adhesion, and then peeled away at a slow rate. This process allowed deterministic transfer of selected thin flakes from PDMS to the silicon surface.\\

\bmhead{THz pump optical probe experiments}
THz pump-optical probe measurements were performed using a custom-built ultrafast spectroscopy setup. The light source was a Ti:sapphire regenerative amplifier (Coherent Astrella) delivering 30\,fs pulses at a repetition rate of 1\,kHz centered at 800\,nm. A portion of the output served as the optical probe, while the remaining beam pumped an optical parametric amplifier (OPA from Light Conversion) to generate 1550\,nm pulses with a duration of approximately 100\,fs. These pulses were focused into a DSTMS crystal to generate single-cycle THz radiation via optical rectification. The THz beam was guided and focused onto the NbOI\textsubscript{2} flakes using a series of off-axis parabolic mirrors, producing a spot size of  $\sim 300\,\mu\mathrm{m}$  on the sample. The probe beam was focused to a spot size of approximately 60\, $\mu\mathrm{m}$ to ensure spatial overlap with the THz pump. The polarization of the THz pump was controlled by adjusting the polarization of the incident optical beam using an achromatic half-wave plate mounted on a motorized rotation stage, in combination with the orientation of the DSTMS crystal. Time resolved polarization rotation measurements were performed in reflection geometry by monitoring the rotation of the linearly polarized 800\,nm probe upon reflection from the NbOI\textsubscript{2} surface. The probe was incident at near-normal incidence, and the reflected beam was analyzed using a balanced detection scheme consisting of a half-wave plate, a Wollaston prism, and a pair of photodiodes. To suppress intensity noise and enhance detection sensitivity, lock-in detection was employed with the pump beam modulated by an optical chopper. \\

\bmhead{Two-dimensional THz spectroscopy}
For the two-dimensional THz pump–probe measurements, the 1550\,nm output of the OPA was split into two beams of equal power to generate a pair of THz excitation pulses. The two beams were subsequently recombined and focused into a DSTMS crystal to produce two single-cycle THz pulses via optical rectification. The relative delay between the two THz pulses was controlled using a mechanical delay stage, defining the interpump delay $\tau$. The THz pulses were then collinearly focused onto the sample. The THz-field-induced optical polarization rotation was detected using the same balanced detection scheme described above, in which the reflected optical probe beam was separated into two orthogonal polarization components and measured with a pair of photodiodes. The time delay between the first THz pulse and the optical probe pulse is defined as \textit{t}. To isolate the nonlinear response, a mechanical chopping scheme was employed to modulate the two THz excitation pulses, denoted A and B. The polarization-rotation signal was obtained by sequentially recording three time-domain traces while scanning the interpulse delay $\tau$ and \textit{t} : the response with both THz pulses present ($S_\text{AB}$), with only pulse A present ($S_\text{A}$), and with only pulse B present ($S_\text{B}$). All measurements were performed under identical experimental conditions. The nonlinear time-domain signal was then extracted by subtracting the linear contributions:
\begin{equation}
S_{\mathrm{NL}}(t,\tau)=S_\text{AB}(t,\tau)-S_\text{A}(t,\tau)-S_\text{B}(t,\tau)
\end{equation}
Finally, a two-dimensional Fourier transform along both the probe delay \textit{t} and the interpump delay $\tau$ was performed to obtain the 2D THz spectrum, which resolves correlations between excitation and detection frequencies and is sensitive to coherent coupling between phonon modes.\\

\bmhead{Device fabrication for in situ electric-field tuning of ferron upconversion}
Two device fabrication strategies were implemented to enable in situ electric-field tuning of ferron upconversion in NbOI\textsubscript{2}. In the first approach, two graphene electrodes were mechanically exfoliated using NITTO SPV224 cleanroom tape and transferred onto a substrate using a polycarbonate (PC) stamp, forming an electrode spacing of 25\,$\mu$m. NbOI\textsubscript{2} was then mechanically exfoliated onto a PDMS stamp, and a flake with a thickness of approximately 150\,nm was identified under an optical microscope. The selected flake was transferred onto the substrate such that its ferroelectric \textit{b}-axis was aligned along the graphene electrode direction. The graphene electrodes were subsequently connected to a breadboard using silver paste and gold wires. To improve fabrication reproducibility, electrodes were also defined using lithographic patterning for subsequent devices. In this process, a bilayer resist stack (MMA 8.5L / PMMA 950A2) was spin-coated onto a Si/SiO\textsubscript{2} substrate and patterned using an Elionix GS-100 electron-beam lithography system. After development in a mixture of isopropanol and deionized water (7:3 ratio), metal electrodes were deposited by thermal evaporation, consisting of 2\,nm Ti as an adhesion layer followed by 10\,nm Pt. The resulting electrode geometry consisted of rectangular pads connected by narrow leads, with an inter-electrode spacing of approximately 50\,$\mu$m. Following electrode fabrication, NbOI\textsubscript{2} flakes were mechanically exfoliated onto PDMS and transferred onto the electrode structure with the polar axis aligned along the electrode direction. The electrodes were wire-bonded using gold wires and silver paste and connected to a DC voltage source (Keithley 2400) to apply the static electric-field bias during measurements.\\

\bmhead{First-principles calculation}

Density functional theory calculations are performed using the Vienna \textit{Ab Initio} Simulation Package (VASP) \cite{VASP1,VASP2,VASP3}. The projector augmented wave (PAW) method \cite{Blochl1994} implemented in VASP is used with a kinetic energy cutoff of 600\,eV. The generalized gradient approximation of Perdew-Burke-Ernzerhof (PBE) \cite{PBE} is used as the exchange-correlation functional. A primitive cell of 8 atoms (2Nb, 4I, and 2O) was used. A $\Gamma$-centered $12\times12\times6$ k-point mesh is used for integration in the first Brillouin zone. Structure is fully relaxed until the force on each atom is less than 0.005\,eV/\AA. DFT-D3 method is used to account for the Van der Waals dispersion correction \cite{grimme_consistent_2010}. Phonon properties are calculated using the Phonopy code \cite{Togo_2023}. A $3\times3\times3$ supercell is used to calculate the interatomic force constants based on the finite difference method with a displacement of 0.01\,\AA. Phonon properties were obtained by diagonalizing the dynamical matrix, which is built on the interatomic force constants. The Born effective charge tensor on each atom is calculated by the density functional perturbation theory implemented in VASP. The ionic contribution to dielectric function is calculated by the following equation~\cite{DFPT}:
\begin{equation}
    \epsilon_{\alpha\beta}(\omega) = \frac{4\pi}{\Omega_0} \sum_m \frac{(\sum_{\kappa\alpha'} Z_{\kappa, \alpha\alpha'}^{*} U_{m\mathbf{q}=0}(\kappa\alpha'))(\sum_{\kappa'\beta'} Z_{\kappa', \beta\beta'}^{*} U_{m\mathbf{q}=0}(\kappa'\beta'))}{\omega_m^2 - \omega^2},
\end{equation}
where $\alpha$ and $\beta$ represent the Cartesian directions $x$, $y$ or $z$, $\kappa$ is the atom index, $Z^*$ is the Born effective charge tensor, $U_m$ is the eigenvector for phonon mode $m$, and $\Omega_0$ is the volume of the unit cell.\\

\section*{Data availability}
The data used in this work are available from the corresponding authors upon reasonable request. \\


\section*{Acknowledgments}
We thank Yujia Zhu, Jiaxuan Wu and Jiamian Hu for fruitful discussions. S.S., F.F., W. F., Y. P. and J.X. acknowledge the primarily support for this research by NSF through the University of Wisconsin Materials Research Science and Engineering Center (DMR-2309000). J. R. and J.X. acknowledge the additional support from from the Office of Naval Research (Grant No. N00014-24-1-2068). Z. Z and B. L. acknowledge the support by U.S. Air Force Office of Scientific Research (grant no. FA9550-19-1-0037), National Science Foundation (grant no. 25163643), and Office of Naval Research (grant no. N00014-23-1-2020). The computational work used the TACC Stampede3 system at the University of Texas at Austin through allocation PHY240212 from the Advanced Cyberinfrastructure Coordination Ecosystem: Services and Support (ACCESS) program~\cite{Boerner2023}, which is supported by US National Science Foundation grants No. 2138259, No. 2138286, No. 2138307, No. 2137603, and No. 2138296.\\

\section*{Author contributions}
S.S. and J.X. conceived the research. S.S., F.F., and J.X. designed the experiments. B. L., Y.P. and J.X. supervised the project. Z.Z. synthesized the bulk high-quality \ce{NbOI2} crystals under the guidance of B.L. . The \ce{NbOI2} samples and the two-terminal \ce{NbOI2} devices were prepared by S.S. with the assistance from J.R., C.F., and A.D.. The ultrafast dynamics measurement was conducted by S.S. with the assistance of F.F. The experimental data was analyzed by S.S., F.F. and J.X.. The first-principles calculations were performed by W.F. under the guidance of Y.P. All authors discussed the results and jointly wrote the paper.\\

\subsection*{Competing interests}
The authors declare no competing interests.\\

\backmatter
\bibliography{sn-bibliography}
\clearpage

\end{document}